\documentclass[twocolumn]{article}
\usepackage{usenix}
\usepackage{times}
\usepackage{scrcode}

\title{\Large \bf Higher-Order Concurrent Win32 Programming}
\author{Riccardo Pucella\\
{\em Bell Laboratories}\\
{\em Lucent Technologies}\\
{\normalsize riccardo@research.bell-labs.com}}
\date{}

\newcommand\kw[1]{\textbf{#1}}

\begin{document}
\maketitle
\thispagestyle{empty}  %

\subsection*{Abstract}
We present a concurrent framework for Win32 programming based on
Concurrent ML, a concurrent language with higher-order functions,
static typing, lightweight threads and synchronous communication
channels. The key points of the framework are the move from an event
loop model to a threaded model for the processing of window
messages, and the decoupling of controls notifications from the system 
messages. This last point allows us to derive a general way of writing 
controls that leads to easy composition, and can accommodate ActiveX
Controls in a transparent way.

\section{Introduction}

Programming user interfaces on the Windows operating system at the
level of the Win32 API, its lowest level, is hard under the best
conditions and maddening the rest of the time. At least three
points underlie the difficulty: 
\begin{itemize}
\item the API is large (more than a thousand functions) and growing;
\item the system is input-driven, making it difficult to perform
intensive computations and remain interactive;
\item the system is centered around an event loop, yielding
control-flow that is hard to understand.
\end{itemize}

The first problem is hard to circumvent, considering that the API is
large because it covers a lot of functionality required by
applications. However, a better structuring can help
reduce the burden of using the API, as demonstrated by the popularity
of C++ class frameworks such as \emph{Microsoft Foundation Classes}
(MFC) or Borland's \emph{Object Windows Library} (OWL). The remaining
two problems are in fact related, consequences of Win32 being a
message-passing API based on callback procedures \cite{Myers91}.

It has been recognized that to ensure interactivity of applications
in the presence of computation-intensive code, multiple threads
should be used \cite{Beveridge96,Gansner92}. It turns out that by using 
a truly concurrent approach rather than the system-level threads
available from the NT kernel, one can construct a framework that also solves the 
last problem, the event loop control-flow nightmare
\cite{Pike89,Haahr90,Gansner92}. 

EXene \cite{Gansner93} is a user interface toolkit for the X
Windows windowing system \cite{Scheifler86}, built on top of
Concurrent ML, a concurrent language providing higher-order functions,
static typing, lightweight threads and synchronous communication
channels \cite{Reppy91}. EXene uses a threading model from the grounds 
up, leading to a design both elegant and simplifying a lot of the
difficulties typically encountered in user interface toolkits. It is
important to note that eXene interacts directly via the X protocol,
without relying on an underlying toolkit.

The goal of this paper is to isolate the difficulties in providing an
eXene-style framework for programming Win32
applications. We focus specifically on the two following points:
moving from an event loop model to a threaded model with
channel-based communication, and decoupling the handling of system
messages from the interaction with controls to help implement easily
composable controls. The resulting system, although 
fairly conservative in its abstractions, is still much simpler to use
than the raw Win32 API, and further supports the thesis that moving 
to a framework based on a high-level concurrent language leads to a
simpler system with good modularity properties.

This paper is structured as follows: after reviewing the structure of
Win32 programs, we describe Concurrent ML and give an outline of the
framework, focusing on the important points of window
management. We then describe how controls are handled, including
predefined controls and custom controls. We also describe how to
handle ActiveX Controls within the same framework in as transparent a
way as possible. We conclude with some discussion of related and
future work. 

We assume the reader has a passing knowledge of higher-order languages 
in general.

\section{Win32 programming}

We review in this section the fundamentals of Win32 programming, at
the level of tutorial books such as \cite{Petzold96}. The Win32 API is 
closely linked with a given program structure. A Win32 program has an
entry point \texttt{WinMain}, whose role is to initialize the
application by creating the various windows making up the
interface. The program then goes in a loop, reading
messages from its message queue and dispatching them to the
appropriate window for processing.

\textbf{Classes}. Every window belongs to a class, which needs to be
registered prior to being used. A class sets the default
icons, cursors, background colors and menu for every window of that
class. A class also contains a pointer to a window procedure,
which is a function invoked every time a message is sent to the
window. Since a window procedure is associated with a given class,
this implies that every window of that class share the same window 
procedure. 

\textbf{Window procedures}. A window procedure is called whenever a
message is sent to an application, either by another window or by the
system. Messages are sent when a window is created, moved, resized or
destroyed, when the mouse moves over the client area of the window,
when the mouse buttons are clicked, when a key is pressed and the
window has focus, when the window needs repainting, when a timer
expires, and so on. Windows provides default handling for all of these 
messages, but an application will want to deal with some of them to
provide its functionality.

\textbf{Child windows}. To simplify the creation of user interfaces,
it is possible to use child windows to subdivide the area of a window
into more manageable components. Each child window has its own window
procedure, thereby encapsulating the behavior of the child window and
allowing a certain amount of abstraction. The child window can decide
to only send a few digested messages back to its parent window, which
can deal with them more easily than otherwise possible. Typical
example of child windows include \emph{controls}, such as buttons,
scrollbars and edit controls. 

That's all there is to Win32 programming, really. Everything else
is concerned with the processing of messages and their
arguments, to do things such as drawing in a window (by
processing the \texttt{WM\_PAINT} message), handling mouse movement
(by processing the \texttt{WM\_MOUSEMOVE} message), handling
keyboard input (by processing the \texttt{WM\_KEYDOWN}
message). Handling controls such as pushbuttons and edit controls is
also done through messages. A pushbutton, for example, notifies its parent
window of an interesting event (e.g. it has been clicked) by sending a
\texttt{WM\_COMMAND} message to the parent, with a code identifying
the control and the notification code as arguments.

\section{Concurrent ML}

The language we use to express our concurrent framework is Concurrent
ML (CML) \cite{Reppy91}, a concurrent extension of the mostly-functional
language Standard ML (SML) \cite{Milner97}. SML provides among other
things higher-order functions, static typing, algebraic datatypes and
polymorphic types. CML is provided as a library-style extension to
SML, with the following (simplified) signature:
\begin{centercode}
\kw{structure} CML : \kw{sig}
  \kw{type} 'a chan
  \kw{type} 'a event
  \kw{type} thread_id
  \kw{val} channel : unit -> 'a chan
  \kw{val} spawn : (unit -> unit) -> thread_id
  \kw{val} sendEvt : 'a chan * 'a -> 'a event
  \kw{val} recvEvt : 'a chan -> 'a event
  \kw{val} wrap : 'a event * ('a -> 'b) -> 'b event
  \kw{val} choose : 'a event list -> 'a event
  \kw{val} sync : 'a event -> 'a
\kw{end}
\end{centercode}

CML is based on the notion of a \emph{thread} which is concurrently
executing thread of control. A function \texttt{spawn} is used to
create a thread that evaluates a given function. Communication between 
threads is done via \emph{channels}. Communication is synchronous:
a send blocks until a receive reads the value on the channel, and vice 
versa. To help design abstract communication protocols, a first-class
notion of \emph{event} is introduced. An event decouples the
communication capability of an operation from its actual execution
(synchronization). Sending a value over a channel is actually a
two-step process: we first create an event that says that the
communication operation to be done will send a value over the channel, 
and when we synchronize on that event, the value is sent over the
channel. Synchronization blocks until the communication is
performed. Basic event constructors include \texttt{sendEvt} and
\texttt{recvEvt} for sending and receiving a value over a
channel. Synchronization is performed by the \texttt{sync}
operation. 

Decoupling definition from synchronization allows for the 
building of combinators to describe more refined communication
mechanisms. For example, given an event, the \texttt{wrap} operation
wraps a function around that event creating a new event that behaves
as follows: when you synchronize on the event, the original event is
synchronized on, and the result of the synchronization is fed to the
function which is evaluated. Given a set of events, you can also
create a new event that 
is a non-deterministic choice over all those events with the
\texttt{choose} operation: when you synchronize on the event, one of
the original events is non-deterministically chosen and synchronized
on. Finally, note that channels and events are  
\emph{polymorphic} over the  carried type (represented by the type
variable \texttt{'a}): a channel of type \texttt{int chan} carries
values of type \texttt{int}, and so on.   

Concurrent ML is currently distributed with the Standard ML of New
Jersey compiler \cite{Appel91}.

\section{The basic framework}

\begin{figure*}[p]
\hrule
\medskip
\begin{code}
  \kw{structure} Run : \kw{sig}
    \kw{type} instance
    \kw{val} doit : (instance -> 'a) -> 'a
  \kw{end}
\end{code}
\hrule
\caption{The \texttt{Run} module}
\label{fig:run}
\end{figure*}

We outline in this section our framework for programming Win32 user
interfaces using the concurrency model provided by CML. The framework
is built on top of a direct binding 
of the Win32 API in SML. An overview of the binding as 
well as examples of use are given in \cite{Pucella99}. The binding
was derived from an IDL description of the API using a tool for
compiling IDL descriptions to SML code interfacing the
described API \cite{UNSTABLE:Pucella99}. This turned out surprisingly
well\footnote{There were some interesting issues raised in providing
the Win32 API  bindings --- both in terms of support of the Win32 API in a
strongly-typed setting, and in terms of the mapping from IDL to
SML --- that may or may not be described in a future article.}
and allowed  us to use code found in tutorial material such as
\cite{Petzold96}. 

For our framework, we build on top of the raw Win32 API some 
layers of abstractions that simplify and abstract away from many
low-level details. We still remain very much in the spirit of 
Win32 however, in the sense that most functions are simply lifted from
the underlying API. Abstractions are mainly concerned with replacing
the event loop by an independent thread and allowing a more
compositional treatment of controls.

The framework aims at supporting more or less the functionality
described in the first volume of \cite{Microsoft93}, along with
various simplifying assumptions. This paper further
simplifies matters for the purpose of presentation, and in order not
to overwhelm the reader with superfluous and confusing details. Note
that we only give the \emph{signature} of the modules in this paper,
that is the type of the operations and values provided by the various
modules. Implementation details are not discussed.

Figure \ref{fig:run} presents the \texttt{Run}
module, which is the main entry point of the framework. The main
function of a program in 
our framework is a function of type 
\begin{centercode}
instance -> 'a
\end{centercode}
taking as argument the instance handle of the program and returning some type 
(exactly which type is returned is unimportant). This function will be 
in charge of creating the various windows of the application, and
calling the message loop of the main window. The function
\texttt{doit} of the \texttt{Run} module invokes the main function,
supplying the application instance handle. 

\begin{figure*}[p]
\hrule
\medskip
\begin{code}
  \kw{structure} Icon : \kw{sig}
    \kw{type} icon
    \kw{val} application : icon
    \kw{val} hand : icon
    \kw{val} question : icon
    \kw{val} exclamation : icon
    \kw{val} asterisk : icon
    \kw{val} load : Run.instance * string -> icon
    \kw{val} draw : DC.hdc * int * int * icon -> unit
  \kw{end}
\end{code}
\hrule
\caption{The \texttt{Icon} module}
\label{fig:icon}
\end{figure*}

\begin{figure*}[p]
\hrule
\medskip
\begin{code}
  \kw{structure} Menu : \kw{sig}
    \kw{type} menu
    \kw{val} load : Run.instance * string -> menu
    \kw{val} get : Window.window -> menu
    \kw{val} create : unit -> menu
    \kw{val} createPopup : unit -> menu
    \kw{val} appendItem : menu * int * string -> unit
    \kw{val} appendPopup : menu * menu * string -> unit
    \kw{val} destroy : menu -> unit
  \kw{end}
\end{code}
\hrule
\caption{The \texttt{Menu} module}
\label{fig:menu}
\end{figure*}

Various modules are provided that simply encapsulate some aspect of 
the API, lifting the functions without trying to generalize or
abstract away some of the functionality. For example, Figure
\ref{fig:icon} and \ref{fig:menu} present modules that deal
respectively with icons and menus. Other modules such as
\texttt{Cursor}, \texttt{Bitmap}, \texttt{Rect}, \texttt{Pen},
\texttt{DC} encapsulate different aspects of the API. All of these are 
fairly straightforward, and aside from their sheer number, their
implementation does not offer any difficulties. It is definitely the
case that future work should aim at finding new abstractions to reduce
either the size or the complexity of this part of the framework.

\begin{figure*}[p]
\hrule
\medskip
\begin{code}
  \kw{structure} Window : \kw{sig}
    \kw{type} class
    \kw{type} window         
    \kw{datatype} class_style = CS_HREDRAW 
                         | CS_VREDRAW 
                         | ...
    \kw{datatype} window_style = WS_OVERLAPPEDWINDOW
                          | ...
    \kw{datatype} show_style = SW_NORMAL
                        | ...
    \kw{val} class : string * Run.instance * Cursor.cursor * Icon.icon * Brush.brush * class_style list -> class
    \kw{val} unregister : class -> unit
    \kw{val} create : class * string * window_style list * window option * int option * int option * int option * 
                 int option * Menu.menu option * Run.instance * (window * Msg.msg chan -> unit) -> window
    \kw{val} createChild : class * string * window_style list * window * int option * int option * 
                      int option * int * int * Run.instance * (window * Msg.msg chan -> unit) -> window
    \kw{val} show : window * show_style -> unit
    \kw{val} update : window -> unit
    \kw{val} setForeground : window -> unit
    \kw{val} move : window * int * int -> unit
    \kw{val} getClientRect : window -> Rect.rect
    \kw{val} destroy : window -> unit
    \kw{val} send : window * Msg.msg -> unit
    \kw{val} quit : int -> unit
    \kw{val} msg_loop : window -> int
    \kw{val} default : window * Msg.msg -> unit
  \kw{end}
\end{code}
\hrule
\caption{The \texttt{Window} module}
\label{fig:window}
\end{figure*}

\begin{figure*}[p]
\hrule
\medskip
\begin{code}
  \kw{structure} Msg : \kw{sig}
    \kw{datatype} msg = WM_SIZE of int * int
                 | WM_PAINT of Rect.rect
                 | WM_DESTROY
                 | WM_TIMER of int
                 | ...
  \kw{end}
\end{code}
\hrule
\caption{The \texttt{Msg} module}
\label{fig:msg}
\end{figure*}

The most important module from our point of view is the one that
focuses on \emph{window management}. Window management describes
anything that relates to the manipulation of windows, including their
creation, deletion, movement, as well as the management of the
classes. As we saw, every window belongs to a class, that assigns
a default icon, cursor and colors for every window of that
class. Moreover, in raw Win32, the class also provides a window
procedure to process messages to the window. The window procedure is
shared amongst all windows of the class. It is not clear why this
design was chosen. Informal explanations are given that this helps 
guarantee that every window of a given class can behave the same
way. But since the window procedure upon reception of a message also
receives the handle of the window to which the message is addressed,
it is very easy to write a window procedure that handles messages
differently depending on the recipient of the message. 

In our framework, we would like to have a thread replacing the window
procedure, and actual messages over channels instead of the Win32
messages passed to window procedures. In order to keep messages
lightweight, we would like to drop the requirement of passing the
handle of the target window when a message is sent. Indeed, our
function to send a message should extract the communication channel
from the window type, and send the message to that channel,
implicitly determining which window the message is sent to. To help
this setup, we will have a thread assigned on a \emph{per window}
basis. Of course, one can still support shared processing amongst all
windows of a given class by delegating every messages to a
centralized thread that communicates with every window of a class.

Figure \ref{fig:window} presents an excerpt of the \texttt{Window}
module containing the interesting parts of the code. Types are defined 
for classes, windows, and various style parameters for both classes
and windows. A function \texttt{class} creates a class given the
appropriate parameters, and automatically registers it. The functions
\texttt{create} and \texttt{createChild} are 
used to create windows, given the class, title, optional owner window, 
position and size (a value of \texttt{NONE} for these forces the use
of a default, equivalent to a \texttt{CW\_USEDEFAULT} in raw Win32),
optional menu, 
instance handle and a function to process messages. This last function 
is spawned automatically on its own thread and is passed the window
being created and a channel to communicate with the window. A child
window is similar, but  
instead of a menu it takes an integer that should uniquely identify the 
child window and that will be use to communicate with the parent
window. Functions are then provided to show, move and destroy the
window. A function \texttt{msg\_loop} is used to initiate the message
loop of a window\footnote{This assumes that windows in the framework
use standard message loops, a simplifying assumption.}. A function
\texttt{send} is used to send a message to a window. The function
\texttt{quit} simply posts the \texttt{WM\_QUIT} message in the
message queue of the application, a requirement for exiting a message
loop. 

As an example, consider the following main function for an application
that bounces a logo around a window. This example is taken from
chapter 7 of \cite{Petzold96}, and is given in its entirety in
Appendix \ref{app:bounce}. It is as simple an initialization function
as can be: only one class, a window created with mostly default
values, and a simple message loop.

\begin{centercode}
\kw{fun} winmain (instance) = \kw{let}
  \kw{val} c = Window.class 
                ("BouncingSMLN", instance, 
                 Cursor.arrow, Icon.application,
                 Brush.white,
                 [Window.CS_HREDRAW, 
                  Window.CS_VREDRAW])
  \kw{val} w = Window.create 
                (c, "Bouncing SML/NJ",
                 [Window.WS_OVERLAPPEDWINDOW],
                 NONE, NONE, NONE, NONE, NONE,
                 NONE, instance, bounce)
  \kw{val} v = Window.msg_loop (w)
\kw{in}
  Window.unregister (c);
  v
\kw{end}
\end{centercode}

The module \texttt{Msg}, outlined in Figure\ref{fig:msg}, defines the
various messages that can be sent to windows by the system and 
by other windows through the \texttt{Window.send} function. There is a
datatype constructor per message, and message parameters are
automatically unfolded for easy retrieval and building. The function
given to \texttt{Window.create} will be spawned and passed the newly
created window and a newly created channel on which the thread will
receive its messages. At 
this point, Win32 rules for processing messages apply: every
message not processed by the application must be passed to default
processing, which means invoking \texttt{Window.default} with the
message as argument, and so on. Often, the thread will simply read
from the input channel and process the messages, but it can also
listen concurrently for events coming from other parts of the
application or from controls.

Finally, although we will not discuss them here, we mention that most
errors in Win32 functions get mapped to SML exceptions.

\section{Controls}
\label{sec:controls}

The first step in the creation of our concurrent framework for Win32
involved lifting window procedures into actual threads with which 
one can communicate using CML-style message-passing. We now turn to the 
second important aspect of our framework: compositional controls. 

A control is ``... a child window an application uses in conjunction
with another window to carry out simple input and output (I/O)
tasks.'' \cite{Microsoft93}. In reality, controls can achieve any
level of complexity chiefly through composition: putting a bunch of
controls together forms a bigger control with potentially a
higher-level semantics. It is possible in raw Win32 to compose
controls, but the amount of plumbing one has to write is
mind-numbing. Our aim  is to make creating new controls by combining
existing ones easy, while staying within the philosophy of Win32. 

A requirement for this to work is that there be no difference between
a predefined control (such as a pushbutton or an edit control) and a
composed control. We also would like the communication to and from the 
control to be independent of the window procedure of the parent
window. The basic idea is that a control
will have a notification channel on which it communicates internal
changes and interesting events. Communication to the control is
achieved by invoking appropriate functions acting on the control.

\subsection{Predefined controls}

\begin{figure*}[p]
\hrule
\medskip
\begin{code}
  \kw{structure} PushButton : \kw{sig}
    \kw{type} push_button
    \kw{datatype} notify_msg = BN_CLICKED 
                        | BN_DOUBLECLICKED
    \kw{val} notifyEvt : push_button -> notify_msg event
    \kw{val} create : string * int * int * int * int * Run.instance -> push_button 
    \kw{val} windowOf : push_button -> Window.window
  \kw{end}
\end{code}
\hrule
\caption{The \texttt{PushButton} module}
\label{fig:pushbutton}
\end{figure*}

\begin{figure*}[p]
\hrule
\medskip
\begin{code}
  \kw{structure} Edit : \kw{sig}
    \kw{type} edit
    \kw{datatype} notify_msg = EN_CHANGE
                        | EN_ERRSPACE
                        | EN_HSCROLL
                        | EN_KILLFOCUS
                        | EN_MAXTEXT
                        | EN_SETFOCUS
                        | EN_UPDATE
                        | EN_VSCROLL
    \kw{val} notifyEvt : edit -> notify_msg event
    \kw{val} getSel : edit -> (int * int)
    \kw{val} setSel : edit * int * int -> unit
    \kw{val} replaceSel : edit * string -> unit
    \kw{val} canUndo : edit -> bool
    \kw{val} emptyUndoBuffer : edit -> unit
    \kw{val} undo : edit -> unit
    \kw{val} create : string * int * int * int * int * Run.instance -> edit
    \kw{val} windowOf : edit -> Window.window
  \kw{end}
\end{code}
\hrule
\caption{The \texttt{Edit} module}
\label{fig:edit}
\end{figure*}

Many controls are predefined in Win32. These include various kind of
buttons (push, check, radio), editing controls, list and combo boxes,
scrollbars, and static controls. Providing them in our framework is
fundamentally a matter of presenting them the right way to the
user. For example, Figure \ref{fig:pushbutton} and \ref{fig:edit}
give the modules implementing respectively pushbuttons and edit
controls. Note the similar format of the modules: both define a type
for the control, a datatype defining the various notification messages 
that the control can report, a CML event that a thread can synchronize 
on to get the notification, functions to communicate with the control, 
a function to create the control, and a function to access the control 
as a normal window, allowing one to apply functions from the
\texttt{Window} module.

The problem with such an interface is that it completely contradicts
the default interface for controls implemented in raw Win32. A
predefined control  
sends notifications directly to its parent window by sending a
\texttt{WM\_COMMAND} message to the window procedure, with its control
ID as an argument and the notification as the other. What we want is to
intercept that message and redirect it onto a CML channel.

One way to achieve this is for the system to to transparently create a
child window 
around the control, which will be the parent of the control, in charge 
of capturing the \texttt{WM\_COMMAND} messages and sending them onto a 
CML channel assigned when the control is created. All very
straightforward, but some work is involved in making sure that all the 
messages sent to the control are communicated to the transparent child
window. For example, applying \texttt{Window.move} to the control should
move the control but also move the transparent child window, and
similarly for resizes and most other window operations.

\subsection{Custom controls}

Custom controls are controls defined by the programmer. To create a
new control, a programmer must determine the appearance of the control 
and its interaction with its subcontrols, if any, and its parent. The 
simplest example of a custom control is a layout control, which
is in charge of maintaining the layout of its subcontrols according to
some constraint criterion. Other more involved controls can include
dozens of subcontrols interacting in a complex way. Dialog boxes can
also be seen as a type of complex control. 

By uniformity, we would like custom controls to respect the informal
specifications given in the previous subsection. Technically, a custom
control is a child window, created via the 
\texttt{Window.createChild} function. The thread associated with the
window, in charge of handling messages to the window, defines the
appearance of the control by handling the \texttt{WM\_PAINT} message,
and so on. Communication with subcontrols is achieved by listening for
the notification events from the subcontrols, concurrently with
handling messages for the window. Similarly, a channel for reporting
notification events for the custom control needs to be allocated.

For example, a new control that encapsulates two pushbuttons
might have a single notification message defined as:
\begin{centercode}
\kw{datatype} notify_msg = CLICKED of int
\end{centercode}
which simply reports which button has been clicked, and a controlling
thread processing messages to the window that also listens to
notification events from the two subcontrols and sends the appropriate
notification when clicks occur (assuming a 
notification channel \texttt{notifyCh}, and pushbuttons \texttt{b1}
and \texttt{b2}): 
\begin{centercode}
...
sync (choose ([wrap (recvEvt (ch), handle_message),
               wrap (PushButton.notifyEvt (b1), 
                       \kw{fn} (PushButton.BN_CLICKED) => 
                             send (notifyCh,CLICKED 1)
                        | _ => ()),
               wrap (PushButton.notifyEvt (b2),
                       \kw{fn} (PushButton.BN_CLICKED) => 
                             send (notifyCh,CLICKED 2)
                        | _ => ())]
...
\end{centercode}

Decoupling the logic of the communication with the subcontrols from
the handling of system messages to the control greatly helps
modularizing the code. Indeed, given a custom control, we could easily
reuse the communication logic for some other control having the same
``behavior'', but maybe a wildly different appearance \cite{Krasner88}.

\subsection{ActiveX Controls}

No discussion of controls would be up-to-date without
mentioning \emph{ActiveX Controls} \cite{Chappell96}. The ActiveX
Controls technology goes back to \emph{Visual Basic Extensions} (VBX),
a mechanism for writing control components for use in the Visual Basic
environment. These were generalized to \emph{OLE Controls} for use in
a general COM-based environment \cite{Rogerson97}. The main problem
with OLE Controls is that they required the programmer to implement a
large number of interfaces that had to be present for the control to
be usable. This did not mix well with the lightweight requirement for
downloadable controls over a network, and so ActiveX Controls were
introduced, fundamentally OLE Controls with looser requirements. 

ActiveX Controls are simply COM objects\footnote{They must also
support self-registration.}, and the support for ActiveX Controls in
any framework is based on the corresponding support for COM objects. An
application that can use ActiveX Controls is called a \emph{control
container}. The functionality of an ActiveX Control is divided into
four parts (from \cite{Chappell96}):
\begin{itemize}
\item providing a user interface;
\item allowing the container to invoke the control's methods;
\item sending events to the container;
\item learning about properties of the container's environment and
allowing the control's properties to be examined and modified.
\end{itemize}

As we discussed in \cite{Pucella99}, calling the methods of a COM
object from SML is fairly easy. It is harder to make the
framework into a control container, because that implies presenting
the whole framework as a COM object with the appropriate interfaces that
ActiveX Controls can access to communicate events. This is not
impossible, but most implementations of SML do not allow this
to be done easily. Given a suitable implementation of such
a capability, it is not hard to see how ActiveX Controls fit in the
above framework. Current work on the SML/NJ runtime system is in part
aimed at solving this particular problem.

\section{Related work}

The idea that concurrency helps in programming user interfaces is not
new. Building on the original work of Squint \cite{Pike89} and Montage
\cite{Haahr90}, eXene \cite{Gansner93} exemplifies the consistent use
of concurrency as a foundation for user interface construction
\cite{Gansner92}. More recently, Haggis \cite{Finne95a}, a functional
framework built on top of a concurrent extension to Haskell, also
demonstrated the usefulness of concurrency in such a context. However,
as opposed to eXene and our approach, the model presented to the user
is strictly sequential --- concurrency is only used internally. 

Compositionality of user interface elements is a requirement for a
programmer-friendly toolkit. Systems such as Tk \cite{Ousterhout94} are
mostly based on the notion that a user interface is a widget (in our
terminology, a control) composed of subwidgets. Building a user
interface is a matter of composing the controls together in a
hierarchical fashion. Tk however uses Tcl as its underlying
language, and because of its lack of large-scale programming
structures, it is not well suited to building large systems
(although some large systems have indeed been built using Tcl/Tk). The
basic ideas underlying compositionality are best presented from 
the point of view of the so-called Model-View-Controller approach, and 
we refer the reader to articles such as \cite{Krasner88} for a deeper
coverage of the issues.

Of course, another closely related system is the Microsoft Foundation 
Classes framework, which provide C++ classes structuring most of the
Win32 API. MFC also allows the definition of methods to handle
messages directly, removing the need to explicitly code up the window
procedure. However, the model is still based on an event loop, and it
is still hard to program computation-intensive applications that
remain interactive. Kernel threads must be used to help manage the
complexity. More experience with our system is required before further 
comparison can be made, especially with respect to the efficiency,
maintainability and reuse possibilities of the code.

\section{Conclusion}

We have described in this paper the design of a simple concurrent
framework for Win32 programming, based on a high-level concurrent
language with lightweight threads. The description we have given is
very much an outline, and indeed even our implementation is
incomplete. We have not talked about color, dialog boxes, keyboard and 
mouse handling, multiple-document interfaces, floating menus, common
dialogs, to name a few. 

The important points about our framework are the move from an event
loop foundation to a threaded model, and a decoupling of the
processing of system messages from the notification messages from
controls. This gives us a chance to derive easily composable
controls. It also gives us a natural way to incorporate ActiveX
Controls transparently into the framework. 

Although the framework does not introduce a great many abstractions
over the underlying Win32 API, the framework is still much easier to
use than a raw Win32 system, and the resulting code more modular,
thereby increasing reusability.

\textbf{Future work}. As we mentioned, the framework is quite
simplistic, and does not go as far as it could go to abstract away
from the underlying system. This was an experiment to try to impose a
concurrent 
communication mechanism onto Win32 that supports an abstract view of
controls decoupled from the window procedure, and nothing else. We
tried to stay as close as possible to the raw Win32 programming
style. Future work is planned in two directions. First, this project
is but a first step in implementing a Win32 interface to Standard ML
of New Jersey. The next step is the design
of a real toolkit that can manage both X-windows and
Windows (and eventually others), with an even more abstract notion of
controls. An investigation into the use of reactive sublanguages
\cite{Halbwachs93a} to express the logic behind the controls
interactions in such a toolkit is also in the works.
Second, we plan to investigate the feasibility of transferring some of 
this work to a C/C++ framework, perhaps at the cost of introducing a
custom version of lightweight threads.

\textbf{Acknowledgments}. Thanks to John Reppy for many discussions
relating to the subject of concurrency in user interfaces that led to
the experiment described in this paper.

\textbf{Availability}. The Standard ML of New Jersey distribution is
available from \texttt{http://cm.bell-labs.com/cm/cs/what/smlnj}, and
information on the framework presented here can be found on the
author's web page at
\texttt{http://cm.bell-labs.com/cm/cs/who/riccardo}. 

{\footnotesize
\bibliography{main,unstable}
}

\onecolumn
\appendix

\section{Bounce example}
\label{app:bounce}

\begin{code}
\kw{fun} bounce (window,ch) = \kw{let}
  \kw{val} timerID = 1
  \kw{val} rate = 20
  \kw{val} moveR = 10
  \kw{val} xTotal = 158
  \kw{val} yTotal = 131
  \kw{val} xRadius = 59
  \kw{val} yRadius = 45
  \kw{fun} onTimer (xS,yS,xC,yC,xM,yM,b) = \kw{let}
    \kw{val} hdc = DC.get (window)
    \kw{val} hdcMem = DC.createCompatible (hdc)
    \kw{val} _ = (Bitmap.select (hdcMem,b);
             DC.bitBlt (hdc,xC - (xTotal div 2), yC - (yTotal div 2), 
                        xTotal, yTotal, hdcMem,0 , 0, DC.SRCCOPY);
             DC.release (window,hdc);
             DC.delete (hdcMem))
    \kw{val} xC' = xC + xM
    \kw{val} yC' = yC + yM
    \kw{val} xM' = \kw{if} (xC' + xRadius >= xS) \kw{orelse} (xC' - xRadius <= 0) \kw{then} ~xM \kw{else} xM
    \kw{val} yM' = \kw{if} (yC' + yRadius >= yS) \kw{orelse} (yC' - yRadius <= 0) \kw{then} ~yM \kw{else} yM
  \kw{in}
    (xS,yS,xC',yC',xM',yM',b)
  \kw{end}
  \kw{fun} computeArgs (x,y,b) = (x,y,x div 2, y div 2, moveR, moveR, b)
  \kw{fun} loop (args \kw{as} (xS,yS,xC,yC,xM,yM,b)) = 
    \kw{case} (recv (ch))
      \kw{of} Msg.WM_SIZE (x,y) => loop (computeArgs (x,y,b))
       | Msg.WM_DESTROY => (Timer.kill (window,timerID);
                            Bitmap.delete (b);
                            Window.quit (window,0))
       | Msg.WM_TIMER (t) => \kw{let} 
             \kw{val} args' = \kw{if} (t=timerID) \kw{then} onTimer (args) \kw{else} args
           \kw{in} loop (args') \kw{end}
       | m => (Window.default (window,m); loop (args))
  \kw{fun} init () = 
    \kw{case} (recv (ch))
      \kw{of} Msg.WM_CREATE => (Timer.set (window,timerIR,rate,NONE);
                           loop (0,0,0,0,0,0,
                                 Bitmap.load ("smlnj.bmp")))
       | m => init ()
\kw{in}
  init ()
\kw{end}

\kw{fun} winmain (instance) = \kw{let}
  \kw{val} c = Window.class ("BouncingSMLN", instance, 
                        Cursor.arrow, Icon.application,
                        Brush.white,
                        [Window.CS_HREDRAW, Window.CS_VREDRAW])
  \kw{val} w = Window.create (c, "Bouncing SML/NJ",
                         [Window.WS_OVERLAPPEDWINDOW],
                         NONE, NONE, NONE, NONE, NONE,
                         NONE, instance, bounce)
  \kw{val} v = Window.msg_loop (w)
\kw{in}
  Window.unregister (c);
  v
\kw{end}
\end{code}

\end{document}